\documentclass{aastex63}

\usepackage{graphicx,epsfig}
\usepackage{dcolumn}
\usepackage{bm}
\usepackage{xcolor}
\usepackage{multibib}
\newcites{main}{References}
\newcites{appendix}{References (Appendix)}
\usepackage{subfigure}
\usepackage{soul}
\usepackage{tcolorbox}
\usepackage{amsmath}
\usepackage{verbatim}
\usepackage[normalem]{ulem}

\newcommand{\be}{\begin{equation}}
\newcommand{\ee}{\end{equation}}
\newcommand{\bse}{\begin{subequations}}
\newcommand{\ese}{\end{subequations}}
\newcommand{\bary}{\begin{eqnarray}}
\newcommand{\eary}{\end{eqnarray}}
\newcommand{\bwt}{\begin{widetext}}
\newcommand{\ewt}{\end{widetext}}

\begin{document}

\title{A two-zone photohadronic interpretation of the EHBL-like behavior of the 2016 multi-TeV flares of 1ES 1959+650}

\author{Sarira Sahu}
\email{sarira@nucleares.unam.mx}
\affiliation{Instituto de Ciencias Nucleares, Universidad Nacional Aut\'onoma de M\'exico, \\
Circuito Exterior, C.U., A. Postal 70-543, 04510 Mexico DF, Mexico}

\author{Carlos E. L\'opez Fort\'in}
\email{carlos.fortin@correo.nucleares.unam.mx}
\affiliation{Instituto de Ciencias Nucleares, Universidad Nacional Aut\'onoma de M\'exico, \\
Circuito Exterior, C.U., A. Postal 70-543, 04510 Mexico DF, Mexico}

\author{Luis H. Castañeda Hernández}
\email{pabcdarioluis@hotmail.com}
\affiliation{Facultad de Ciencias, Universidad Nacional Aut\'onoma de M\'exico, \\
Circuito Exterior, C.U., A. Postal 70-543, 04510 Mexico DF, Mexico}

\author{Subhash Rajpoot}
\email{Subhash.Rajpoot@csulb.edu}
\affiliation{Department of Physics and Astronomy, California State University,\\ 
1250 Bellflower Boulevard, Long Beach, CA 90840, USA}

\begin{abstract}
The high-energy-peaked blazar 1ES 1959+650 is a well-known and well studied nearby blazar that has undergone several episodes of multi-TeV flaring. In 2002 for the first time an orphan TeV flare was observed from this blazar. During a multiwavelength campaign between 29th April to 21st November 2016, MAGIC telescopes observed multi-TeV flarings during the nights of 13th, 14th June and 1st July 2016 when the position of the synchrotron peak was found to be above $10^{17}$ Hz. Also observed was that the second peak of the spectral energy distribution shifted towards higher energy, and exhibiting extreme HBL-like behavior. The photohadronic model which is very successful in explaning the multi-TeV flaring from many high energy blazars including 1ES 1959+650 is applied to study the flaring events of 2016. It is observed that the photohadronic model is unable to explain the observed spectra. Here, we use a two-zone photohadronic model to explain the observed spectra. We clearly demonstrate that the low energy regime (zone-1) of the spectra corresponds to the standard flaring events of the high energy blazar and the high energy regime (zone-2) of the spectra are solely due to the extreme nature of the flaring events. Our two-zone photohadronic model explains very well the multi-TeV flaring events observed by MAGIC telescopes.  
\end{abstract}


\section{Introduction}
Blazars 
are the dominant extra galactic population in $\gamma$-rays 
and show rapid variability in the entire
electromagnetic spectrum \citep{Acciari:2010aa}. Their non-thermal spectra are produced
by the relativistic jet which lies close to the
observers line of sight  \citep{Abdo:2010rw}. 
The spectral energy
distribution (SED) of these blazars is characterized by two non-thermal peaks. 
The low-energy peak is from the synchrotron radiation of the relativistic
electrons in the jet and the high-energy peak is from the synchrotron 
self Compton (SSC) scattering of the 
high-energy electrons with their self-produced synchrotron
photons  \citep{Blazejowski:2000ck,Murase:2011cy,Gao:2012sq}. Depending on the energy of their synchrotron peak, they are classified as: low-energy peaked blazars (LBLs, $\nu_{peak} < 10^{14}\ \mathrm{Hz}$), intermediate-energy peaked blazars (IBLs, $10^{14}\ \mathrm{Hz} < \nu_{peak} < 10^{15}\ \mathrm{Hz}$), high energy-peaked blazars (HBLs, $10^{15}\ \mathrm{Hz} < \nu_{peak} < 10^{17}\ \mathrm{Hz}$), and extreme high energy-peaked blazars (EHBLs, $\nu_{peak}>10^{17}\ \mathrm{Hz}$) \citep{Abdo:2009iq,Costamante:2017xqg}.

The 1ES 1959+650 is a well-known HBL at a redshift of $z=0.048$ \citep{perlman:1996aj}. It was first detected in the radio range by NRAO Green Bank Telescope \citep{gregory:1991aj} and in X-ray range by Einstein IPC Slew Survey (Elvis et al. 1992). At TeV energies it was first detected by the Utah Seven Telescope array \citep{Nishiyama:1999js} during an observation campaign in 1998 with a significance of $3.9\sigma$ \citep{Nishiyama:1999js}. In 2001 the HEGRA collaboration reported another TeV outburst with a significance $>4\sigma$ \citep{Horan:2002vd}. 

In May 2002, a strong TeV flare was observed by Whipple and HEGRA experiments from 1ES 1959+650 \citep{Holder:2002ru,Aharonian:2003be}. It was also observed by RXTE experiments in X-rays. However, the X-ray flux smoothly declined throughout the following month and on 4th of June, a second TeV flare was observed without an X-ray counterpart. As this flaring event was not accompanied by any low energy counterparts, it is referred to as orphan flare and is in striking disagreement with the predictions of the leptonic models, thus contradicting the leptonic process as the origin of the TeV emission mechanism. This particular flaring event calls for alternative mechanisms for very high energy (VHE $> 100$ GeV) $\gamma$-ray production in blazars. In one scenario, a hadronic mirror model was proposed to explain this particular event \citep{Bottcher:2004qs}. In such a model  the TeV gamma-rays are produced from the interaction of high-energy protons with the primary synchrotron photons that have been reflected off clouds located at a few pc above the accretion disk. In another scenario, \cite{Sahu:2013ixa} instead proposed a photohadronic model where the $\gamma$-rays are produced from the interaction of Fermi-accelerated protons with the background seed photons in the jet environment.

During August 2004, VHE $\gamma$-ray signals were detected by GT-48 Telescope of CAO \citep{Fidelis:2006cao} and also in September-October by MAGIC telescopes\citep{Albert:2005jv}. In May 2006 several instruments reported a high state in the optical-UV and X-ray bands but low state in VHE $\gamma$-rays, and strong variability in the high X-ray band, suggesting a "standing shock" scenario \citep{Albert:2008uda}. Furthermore, several states of enhanced VHE $\gamma$-ray activities were reported in April, June, and October of 2015 by Fermi-LAT, VERITAS, and other collaborations \citep{Kaur:2017nxb,Acciari:2020spw}. Previously, several of these flaring events have been explained by leptonic and hadronic models \citep{Holder:2002ru,Aharonian:2003be,Bottcher:2004qs,Krawczynski:2003fq,Daniel:2005rv,Gutierrez:2006ak,Sahu:2019kfd}. 

In the year 2016, observation of enhanced flux from optical to $\gamma$-ray from the HBL 1ES 1959+650 triggered extensive observations by MAGIC telescopes \citep{Hayashida:2020wez} and its effective observation window was between 29 April to 29 November.
The highest fluxes in VHE were reported during the nights of June 13th, 14th and July 1st, and no optical and X-ray counterparts were reported on 1st July. 
During this period, it was observed that the synchrotron peak was shifted above $10^{17}$ Hz and also the SSC peak shifted towards higher energy. This shift in synchrotron peak above $10^{17}$ Hz is classified as EHBL and so far many such peaks have been observed \citep{Costamante:2017xqg}. 
A similar shift in the spectra were observed in VHE flaring of Markarian 501 (Mrk 501) in 2005 and 2012 \citep{Albert:2007zd,Ahnen:2018mtr} which are transitions from HBL to EHBL, although it is a well known HBL. 
Thus, it is understood that 1ES 1959+650 underwent a similar transition from HBL to EHBL. 

Recently, we have shown that, the standard photohadronic scenario is inadequate to explain the EHBL-nature of Mrk 501. As in the case of multi-TeV flaring from HBLs, it was observed that the background seed photons in the SSC band participating as the target for the Fermi-accelerated protons have a single power-law behavior given by $\Phi_{SSC} \propto \epsilon^{\beta}_{\gamma}$, with $0.5\le \beta\le 1$. However, for the EHBL case we observed that, the single power-law behavior of the seed photon flux is no longer followed. Thus, by assuming two different power-laws for the SSC band, we have shown that, the extreme flaring of Mrk 501 during 2005 and 2012 can be explained very well \citep{Sahu:2020tko}. It is clearly shown that, the low energy regime (zone-1) is the usual HBL flaring with no change in its flux and spectral index. However, the extreme nature of the spectrum is attributed to the change in the high energy regime (zone-2) which is solely due to the change in the power-law behavior of the seed photon flux in the lower tail region of the SSC spectrum. This change in the seed photon spectral index is responsible for the overall spectral index of the VHE $\gamma$-ray spectrum.

Our main motivation here is to use the two-zone photohadronic model, which successfully explained the EHBL nature of Mrk 501 VHE flaring events, to analyze the extreme nature of HBL 1ES 1959+650 during the three nights of June 13th, 14th and July 1st of 2016.

\section{Two-zone photohadronic model}

The multi-TeV flaring from many HBLs are successfully explained using the photohadronic model \citep{Sahu:2019lwj,Sahu:2019kfd}. During the GeV-TeV flaring epoch, the photohadronic model assumes the formation of a double jet structure where a hidden inner jet of size $R'_f$ and photon density $n'_{\gamma,f}$ is surrounded by an outer jet of size $R'_b$ and photon density $n'_{\gamma}$, with $R'_f<R'_b$ and $n'_{\gamma,f}>n'_{\gamma}$ (where $'$ implies comoving frame). Both the jets are moving with a bulk Lorentz factor $\Gamma$ and Doppler factor $\mathcal{D}$.
The photon density of inner jet decreases as it crosses into the outer jet due to its adiabatic expansion. Also, the photon density in the hidden inner jet is unknown. However, we assume a scaling behavior \citep{Sahu:2019lwj} to relate the photon density in the inner and the outer region. Mathematically this is given as,
\be
\frac{n'_{\gamma,f}(\epsilon_{\gamma,1})}{n'_{\gamma,f}(\epsilon_{\gamma,2})} \simeq\frac{n'_{\gamma}(\epsilon_{\gamma,1})}{n'_{\gamma}(\epsilon_{\gamma,2})}.
\label{eq:scaling}
\ee
The above relation shows that, during the flaring episode of the HBL,
the ratio of the photon densities at energies $\epsilon_{\gamma,1}$ and  $\epsilon_{\gamma,2}$ in the inner jet and the outer jet region are almost the same. From the observed flux, the photon density in the outer region can be calculated and by using the relation in Eq. (\ref{eq:scaling}), we can express the inner photon density in terms of the observed flux. The accelerated high energy protons in the inner jet interact with the low-energy background seed photons in the SSC region through $p\gamma\rightarrow\Delta^+$ process. Subsequently the $\Delta$-resonance decays into $\gamma$-rays through intermediate $\pi^0$ and to neutrinos via $\pi^+$.

In the inner jet, the injected proton spectrum is a power-law in its energy $E_p$ \citep{Dermer:1993cz} and expressed as
$dN/dE_p\propto E^{-\alpha}_p$,
where the spectral index $\alpha \ge 2$. The kinematical condition \citep{Sahu:2019lwj} to produce $\Delta$-resonance is, 
\be
E_p \epsilon_\gamma=\frac{0.32\ \Gamma\mathcal{D}}{(1+z)^{2}}\ \mathrm{GeV^2},
\label{eq:kinproton}
\ee
where $\epsilon_\gamma$ is the seed photon energy and $z$ is the redshift of the object. For HBLs, $\Gamma \approx \mathcal{D}$ and the observed VHE $\gamma$-ray carries about $10\%$ of the proton energy i.e $E_{\gamma}\simeq 0.1\, E_p$. So far, in all the GeV-TeV flaring of HBLs, we have observed that the range of $E_{\gamma}$ corresponds to the seed photon energy $\epsilon_{\gamma}$ in the low energy tail region of the SSC spectrum and in this region the SSC flux is a perfect power-law, given as $\Phi_{SSC}\propto \epsilon^{\beta}_{\gamma}$ \citep{Sahu:2019lwj}. 

The EBL plays a very important role in the attenuation of VHE gamma-rays and its effect should be accounted for in the observation of VHE gamma-rays \citep{Padovani:2017zpf}. A number of EBL models have been developed using different methods \citep{Salamon:1997ac,Stecker:2005qs,Franceschini:2008tp,Dominguez:2010bv,Gilmore:2011ks,Dominguez:2013lfa}. 
By accounting for the EBL correction in the photohadronic model , the observed multi-TeV spectrum can be given as 
\be
F_{\gamma,obs}(E_{\gamma})=F_0 \left ( \frac{E_\gamma}{TeV} \right )^{-\delta+3}\,e^{-\tau_{\gamma\gamma}(E_\gamma,z)}=F_{\gamma, in}(E_{\gamma})\, e^{-\tau_{\gamma\gamma}(E_\gamma,z)},
\label{eq:fluxgeneral}
\ee
where $F_0$ is the normalization constant and can be fixed from the observed VHE spectrum. The exponent $\delta=\alpha+\beta$ is the only free parameter in this model. $F_{\gamma,in}$ is the intrinsic VHE gamma-ray flux. Previously we have studied many flaring HBLs and observed that $\delta$ lies in the range $2.5\le \delta \le 3.0$ and depending on the value of $\delta$, the flaring states can roughly be classified into three categories as: (i) very high state with $2.5\le\delta\le 2.6$, (ii) high state with $2.6<\delta<3.0$ and (iii) low state for $\delta=3.0$ \citep{Sahu:2019kfd}. In general, the proton spectral index $\alpha=2$ is taken. So, once the value of $\delta$ is fixed by fitting the VHE spectrum, we can calculate the value of $\beta$ and compare with the leptonic model fit.

Mrk 501, is a well known and extensively studied HBL whose VHE spectra are well explained by the photohadronic model \citep{Sahu:2019lwj,Sahu:2019scf}. However, during 2005 and 2012 it had undergone transition from HBL to EHBL state \citep{Albert:2007zd,Ahnen:2018mtr}. Recently, we have shown that these extreme nature of the flaring events do not fit well when applied to the above photohadronic model. Instead, we observed that a two-zone photohadronic scenario fares much better when applied to explain these events \citep{Sahu:2020tko}. In this case, it is assumed that the background seed photon flux in the tail region of the SSC band can be expressed by two power-laws as, 
\be
\Phi_{SSC}\propto
 \left\{ 
\begin{array}{cr}
E^{-\beta_1}_{\gamma}
, & \quad 
 \mathrm{100\, GeV\, \lesssim E_{\gamma} \lesssim E^{intd}_{\gamma}}
\\ E^{-\beta_2}_{\gamma} ,
& \quad   \mathrm{E_{\gamma}\gtrsim E^{intd}_{\gamma}}
\\
\end{array} \right. ,
\label{eq:sscflux}
\ee
where $\beta_1\neq \beta_2$, $E^{intd}_{\gamma}$ is an energy scale around which the transition between zone-1 and zone-2 takes place whose value can be fixed from the individual flaring spectrum. Using the above SSC flux the observed VHE spectrum can be expressed as
\be
F_{\gamma, obs}=
e^{-\tau_{\gamma\gamma}}\times
\begin{cases}
 F_1 \, \left ( \frac{E_{\gamma}}{TeV} \right )^{-\delta_1+3}
, & \quad 
\mathrm{100\, GeV\, \lesssim E_{\gamma} \lesssim E^{intd}_{\gamma}}\,\,\,\,\, (\text{zone-1}) \\ 
F_2 \, \left ( \frac{E_{\gamma}}{TeV} \right )^{-\delta_2+3},
& \quad \,\,\,\,\,\,\,\,\,\,\,\,\,\,\,\,\,\,\,\,\,\,\,\,\,\,\,\,\,\,\, \mathrm{E_{\gamma}\gtrsim E^{intd}_{\gamma}}\,\,\,\,\, (\text{zone-2})
 \end{cases}.
\label{eq:flux}
\ee
Here, $F_1$ and $F_2$ are normalization constants and the spectral indices $\delta_i=\alpha+\beta_i$ ($i=1,2$) are the free parameters to be adjusted by fitting to the observed VHE spectrum of the EHBL.

\section{EHBL-like multi-TeV flaring events}

In 2016, observation of high flux states from optical to $\gamma$-ray range from the HBL 1ES 1959+650 triggered extensive observations by MAGIC telescopes between 29 April to 29 November, a total observation time of 72 hours spread over 67 nights \citep{Hayashida:2020wez}. These observations included both high and low flux states from the source. The highest fluxes in VHE were reported during the nights of June 13th, 14th and July 1st. Also reported were that no optical and X-ray counterparts were  detected on 1st July. However, intra-night variabilities were observed on June 13th and July 1st nights. During this period, the synchrotron and the SSC spectra were also shifted towards higher energies with the synchrotron peak located above $10^{17}\ \mathrm{Hz}$.
 
\begin{figure}[h]
{\centering
\resizebox*{0.8\textwidth}{0.5\textheight}
{\includegraphics{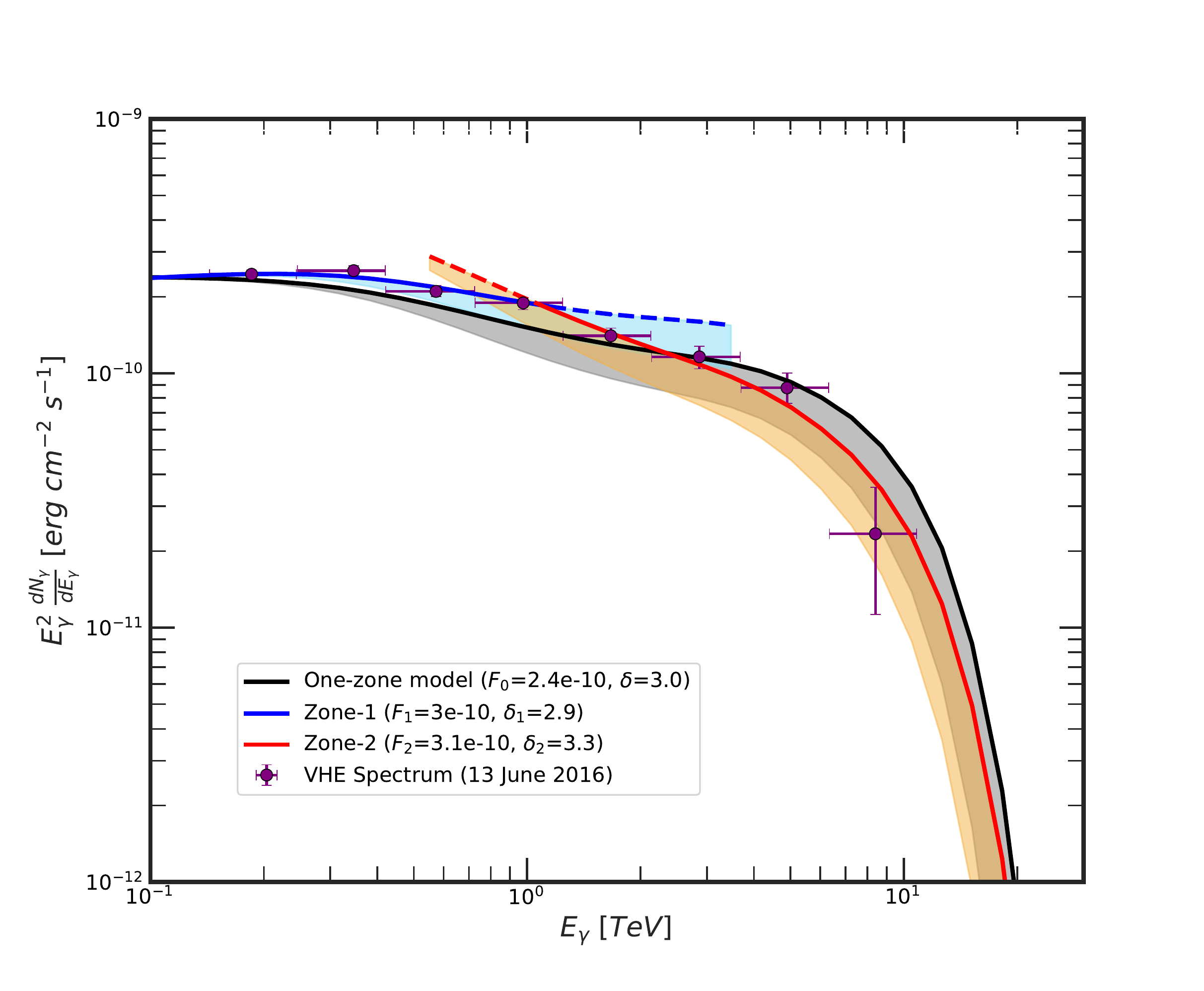}}
\par}
\caption{
The VHE spectrum of 1ES 1959+650 observed in the night of  13th June 2016 \citep{Hayashida:2020wez} is fitted with 
standard photohadronic model (black curve) and also with two-zone photohadronic model. The best fit to zone-1 is shown in pink curve with a statistical goodness of 97.4\% and the dashed pink curve is the behavior of the model in the high energy limit. The spectrum in zone-2 is fitted with blue curve with a statistical significance of 94.6\% and the dashed blue curve shows its behavior in the low energy limit. The shaded regions represent the lower bounds for the VHE $\gamma$-ray flux assuming up to 50\% increase in the optical depth (see Section 3). In Figures \ref{fig:figure2} and \ref{fig:figure3}, the dashed curves and shaded regions have the same interpretation as this figure. In all the figures the normalization constants $F_i (i=0,1,2)$ are expressed in units of $\mathrm{erg\, cm
^{-2}\, s^{-1}}$.
}
\label{fig:figure1}
\end{figure}
\begin{figure}[h]
{\centering
\resizebox*{0.8\textwidth}{0.5\textheight}
{\includegraphics{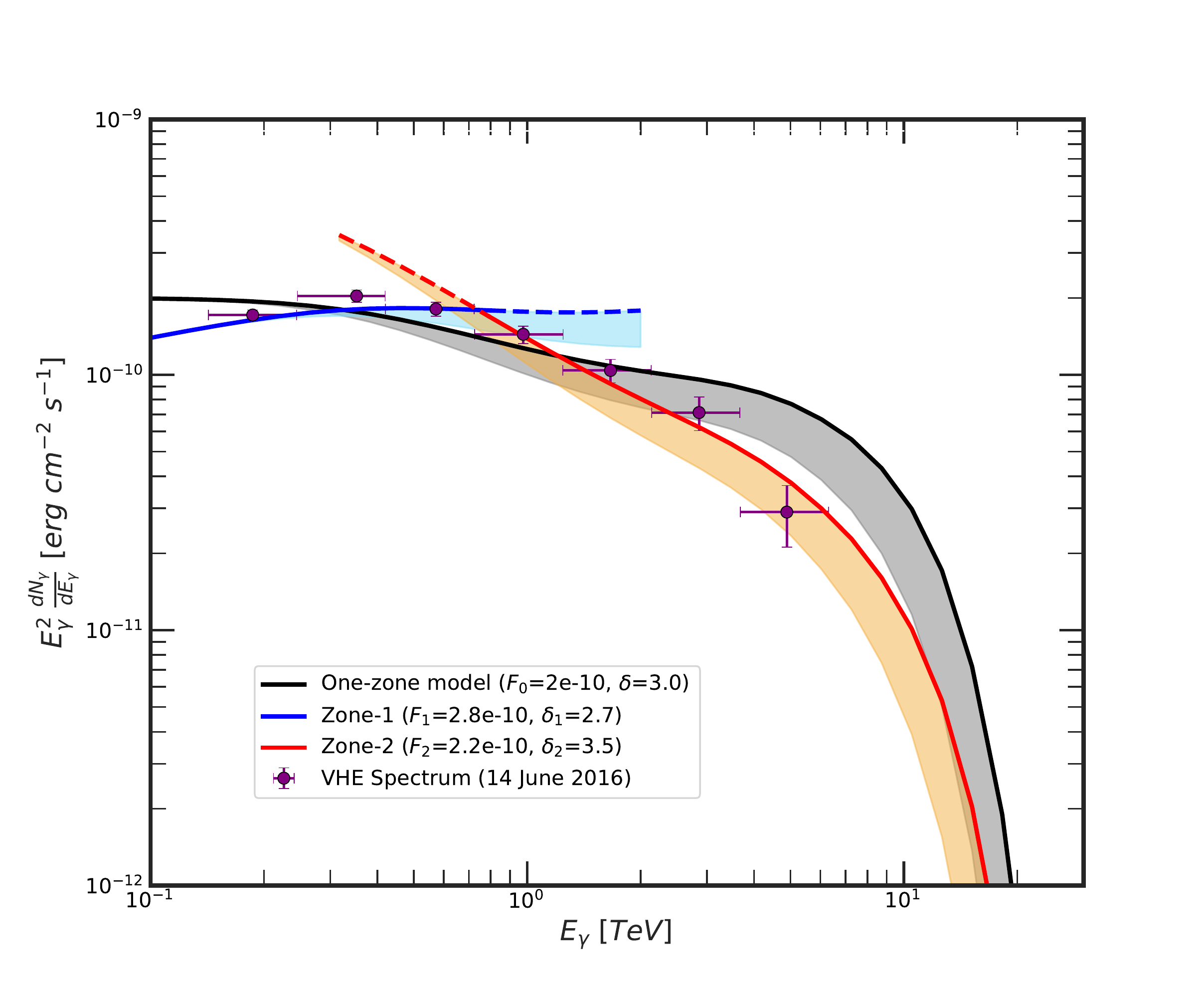}}
\par}
\caption{
The VHE spectrum of 1ES 1959+650 observed in the night of  14th June 2016 \citep{Hayashida:2020wez}. Different curves are having the same interpretation as shown in Figure \ref{fig:figure1}.
}
\label{fig:figure2}
\end{figure}
\begin{figure}[h]
{\centering
\resizebox*{0.8\textwidth}{0.5\textheight}
{\includegraphics{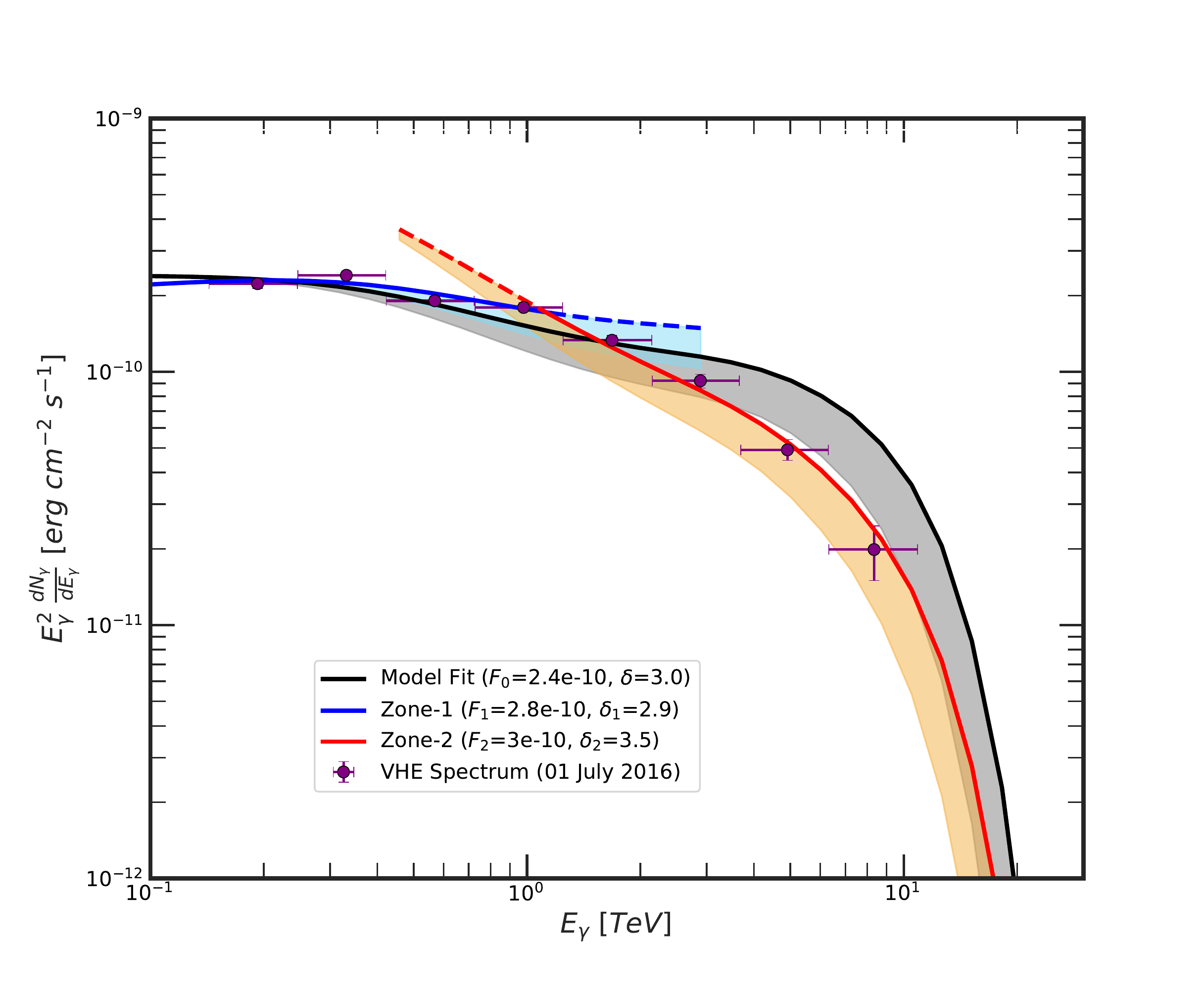}}
\par}
\caption{
The VHE spectrum of 1ES 1959+650 observed in the night of  1st July 2016 \citep{Hayashida:2020wez}. Different curves are having the same interpretation as shown in Figure \ref{fig:figure1}.
}
\label{fig:figure3}
\end{figure}

Previously, the multi-TeV flaring from 1ES 1959+650 during May 2002, November 2007 - October 2013, May 21-27 2006 and May 2012 were explained very well by the photohadronic model \citep{Sahu:2019lwj,Sahu:2019kfd}. So, naturally as a first option, we use this model again to explain the extreme behavior of June 13th, 14th and July 1st, 2016 flaring events. For our analysis, we use ${\cal D}\simeq \Gamma=20$, the inner blob radius during the flaring is assumed to be $R'_f\sim 5\times 10^{15}$ cm and many leptonic models assume $R'_b \gtrsim 10^{16}$ cm \citep{Albert:2008uda,Aliu:2013nza,Kaur:2017nxb}.
Using a standard optimization of parameters from the weighted least squares (WLS) function, we fitted the observed VHE spectra.
The goodness of the fits were tested using a Pearson's $\chi^2$-test with respect to the central values. The uncertainty in the EBL is neglected for simplicity, and the percentages were obtained using a standard $\chi^2$ distribution for $n$ degrees of freedom (were $n$ is equal to number of observed data points minus number of free parameters).  
The best fit to the data are shown in the black curves of Figures \ref{fig:figure1}, \ref{fig:figure2} and \ref{fig:figure3} for the above three nights. However, it is clearly seen that we are unable to fit any of these flaring events in a satisfactory manner. While the photohadronic model fit is little flat with a slight dip between $\sim\,800$ GeV to $\sim\,3$ TeV energy range, the observed spectra are decaying faster and smoother than expected with the EBL correction. We observed that the flattening of the photohadronic fit is due to the EBL effect.

Henceforth, we instead proceed to fit the VHE spectra using the two-zone photohadronic model given in Eq. (\ref{eq:flux}). We assume the same jet parameters as before  and also perform the same optimization to derive the best fits for each zone. We discuss now the VHE emission from individual nights using this modified model. It is important to mention that the VHE data shown in the SED modeling of the MAGIC collaboration in Figs. 6-8 for June 13th and 14th do not correspond to their respective observed spectrum but to the de-absorbed one when compared with Figure 4 \citep{Hayashida:2020wez}.

For our the present analysis, the EBL model of Franceschini et al. \citep{Franceschini:2008tp} is used, which is consistent with a EBL peak flux of $\sim 15\ \mathrm{nW}\mathrm{n^{-2}}\ \mathrm{sr^{-1}}$ at $1.4\, \mathrm{\mu m}$ as constrained by the $\gamma$-ray observations from H.E.S.S \citep{Abramowski:2012ry}. However, recently CIBER, IRTS, and AKARI experiments have reported a higher value of the flux $\sim 28.7\ \mathrm{nW}\  \mathrm{n^{-2}}$ at $1.4\, \mathrm{\mu m}$ exceeding almost two times the previous EBL upper limits \citep{Matsuura:2017lub}. Also, some other hadronic models propose that the observed TeV $\gamma$-ray spectra from blazars and the lack of absorption features could be explained by the production of secondary $\gamma$-rays \citep{Aharonian:2012fu,Essey:2009zg,Essey:2009ju,Essey:2010er}. It should be noted that the VHE flarings of 1ES 0229+200, 1ES 0347+121, and 1ES 1101+232 studied by Essey et al. \citep{Essey:2010er} have also been studied by using the standard photohadronic scenario \citep{Sahu:2019kfd} and using the EBL model of Franceschini et al. \citep{Franceschini:2008tp}.

We can take into account the potential effect of a stronger EBL attenuation as obtained by \cite{Matsuura:2017lub} by assuming that the EBL SED of Franceschini et al. \citep{Franceschini:2008tp} has the same shape but with an uncertain normalization. Adjusting for the CIBER observations in Figure 11 from \cite{Matsuura:2017lub}, we find that this roughly equals to a 50\% increase in the flux of the EBL SED of Franceschini et al. \citep{Franceschini:2008tp}. As the optical depth is proportional to flux, this translates to a maximum increase of the optical depth up to 50\%.

\subsection{June 13th}

On 13th June, MAGIC telescopes observed 1ES 1959+650 for $\sim$ 1.5 h, and 
simultaneously it was also observed in UV/optical and X-ray by UVOT and by XRT-Swift telescopes, respectively. Using leptonic, hadronic and lepto-hadronic models, the multiwavelength SED was reconstructed \citep{Hayashida:2020wez}. Significant intra-night variability above 300 GeV was also reported by MAGIC collaboration. The time-averaged VHE spectrum is in the energy range $0.18\, TeV \le E_{\gamma}\le 8.4$ TeV which corresponds to Fermi accelerated proton energy in the range $1.8\, TeV \le E_{p}\le 84.0$ TeV and the seed photon energy in the range $1.4\, MeV\le \epsilon_\gamma\le 64.7$ MeV. 

The low energy part of the VHE spectrum (zone-1) with $E^{intd}_{\gamma} \lesssim 1$ TeV can be fitted very well with $F_1=3.0\times 10^{-10}\ \mathrm{erg\,cm^{-2}\,s^{-1}}$ and $\delta_1=2.9$, with a statistical significance of 97.4\% and this value of $\delta_1$ corresponds to high emission state. Similarly, the best fit to the high energy part (zone-2) is obtained for $F_2=3.1\times 10^{-10}\ \mathrm{erg\,cm^{-2}\,s^{-1}}$ and $\delta_2=3.3$, with a statistical significance of 94.6\%. These are shown in Figure \ref{fig:figure1} and compared with the standard photohadronic scenario. In zone-1, the seed photon in the background has $\beta_1=0.9$ and in zone-2 $\beta_2=1.3$ as we take $\alpha=2.0$ here. Thus these different spectral behavior of the background seed photons in the SSC spectrum are responsible for two separate zones. To account for the enhanced EBL contribution we have increased the optical depth up to 50\% which apparently decreases the value of $e^{-\tau_{\gamma\gamma}}$ and consequently the observed VHE spectrum. These are shown in the shaded regions of Figure \ref{fig:figure1} for the standard and two-zone photohadronic models. By adjusting the values of $F_1$ and $F_2$ we can fit the observed VHE spectrum with the lower bound of the shaded regions.

To produce $\Delta$-resonance within the inner jet, the optical depth for the $p\gamma$ process
should satisfy $\tau_{p\gamma}<1$ and this constrains the SSC photon density in the region to be $n'_{\gamma,f}< 4.0\times 10^{11} \ \mathrm{cm^{-3}}$. However, we consider a moderate efficiency of the $p\gamma$ process by taking $\tau_{p\gamma}\sim 0.2$ which gives $n'_{\gamma,f} \sim 8.0\times 10^{10} \ \mathrm{cm^{-3}}$.

Considering the central black hole mass $M_{\mathrm{BH}}\sim 3\times 10^{8}M_{\odot}$ \citep{Wagner:2007kp}, the Eddington luminosity of 1ES 1959+650 is $L_{Edd}\sim 3.8\times 10^{46}\ \mathrm{erg}\ \mathrm{s^{-1}}$. The highest energy $\gamma$-ray with $E_\gamma=8.4\ \mathrm{TeV}$ has the flux  $F_\gamma\sim 2.3\times 10^{-11}\ \mathrm{erg}\ \mathrm{cm^{-2}}\ \mathrm{s^{-1}}$ which corresponds to a luminosity of $L_\gamma\sim 1.4\times 10^{44}\ \mathrm{erg}\ \mathrm{s^{-1}}$.
The proton luminosity to produce these $\gamma$-rays is given by $L_p=7.5 L_\gamma \tau_{p\gamma}^{-1}$ and again this should satisfy $L_p<L_{Edd}/2$ which gives $\tau_{p\gamma} > 0.06$. Thus by taking $\tau_{p\gamma}\sim 0.2$, we get $L_p\sim 5.2\times 10^{45}\ \mathrm{erg}\ \mathrm{s^{-1}}$ which is smaller than half of the Eddington luminosity.
 
By considering the multiwavelength SED of June 13th from \cite{Hayashida:2020wez}, and assuming that its minimum SSC energy $\epsilon_\gamma\sim 4\times10^{21}\ \mathrm{Hz}$ is the seed to produce the highest energy $\gamma$-ray of energy $E_\gamma\simeq 8.4$ TeV, we get $\Gamma\sim 21$. This agrees perfectly with the value used in our calculation.

\subsection{June 14th}

On 14th June, the MAGIC telescopes observed for $\sim$ 2 h when the flux above 300 GeV reached up to 3 Crab units. Simultaneous observations in the UV/optical and X-ray ranges were recorded by the UVOT and XRT-Swift telescopes, respectively. No significant intra-night variability was observed on this night and the VHE spectrum observed is in the energy range $0.19\le E_\gamma\le 4.9$ TeV \citep{Hayashida:2020wez}. The low energy part of the spectrum with $E^{intd}_{\gamma} \lesssim 1$ TeV corresponds to zone-1 and is best fitted with $F_1=2.4\times 10^{-10}\ \mathrm{erg\,cm^{-2}\,s^{-1}}$ and $\delta_1=2.9$, with a statistical significance of 92.6\%. This is also in a high state. The zone-2 with $E^{intd}_{\gamma} \gtrsim 1$ TeV is well fitted with $F_2=2.3\times 10^{-10}\ \mathrm{erg\,cm^{-2}\,s^{-1}}$ and $\delta_2=3.5$, which has a 92.3\% statistical significance. The two-zone and standard photohadronic models are shown in Figure \ref{fig:figure2} for comparison. The fits to two-zone models corresponds to $\beta_1=0.9$ and $\beta_2=1.5$,  respectively. 
The efficiency for $\Delta$-resonance production and the seed photon density in the jet is very similar to that of the previous night estimate. The shaded regions in Figure \ref{fig:figure2} correspond to an increase up to 50\% in optical depth which is very similar to the one discussed in Section 3.1.

\subsection{July 1st}
Unlike the previous nights, on July 1st, no simultaneous observations in the UV/optical and X-ray ranges were registered by the UVOT and XRT-Swift telescopes. 
However, it presented high intra-night variability over short timescales and the 
observed VHE spectrum is in the energy range 
$0.19\le E_\gamma\le 8.3$ TeV \citep{Hayashida:2020wez}. This corresponds to proton energy in the range $2\, TeV\le E_p\le 83$ TeV and the seed photon in the energy range   
$1.4\le \epsilon_\gamma\le 61.4$ MeV respectively. The best fit to the observed spectrum in zone-1 is obtained for $F_1= \mathrm{2.8\times 10^{-10} \ erg \ cm^{-2}\ s^{-1}}$, and $\delta_1=2.9$, with a statistical significance for the goodness of the fit of 94.2\%. This value of $\delta$ implies that the flaring was in high emission state. The zone-2 is also fitted extremely well with $F_2= \mathrm{3.0\times 10^{-10} \ erg \ cm^{-2}\ s^{-1}}$, and $\delta_2=3.5$, with a statistical significance of 98.3\%.
As before, in Figure \ref{fig:figure3} we have shown the observed data, the photohadronic fit and the two-zone photohadronic fit. Again, one can clearly see that the standard photohadronic model is unable to fit the observed data. In this case also, the seed photon density and $p\gamma$ optical depth are similar to that of the previous two nights. The shaded regions in Figure \ref{fig:figure3} correspond to an increase up to 50\% in optical depth, with a similar interpretation to the one discussed in Section 3.1.

\section{Discussion}

Since its discovery in high-energy, the HBL 1ES 1959+650 had undergone many episodes of flaring in multi-TeV energy. Also in the year 2002 it had a orphan flaring in VHE gamma-rays. Most of these flaring events were explained very well using the standard photohadronic model \citep{Sahu:2019lwj,Sahu:2019kfd}. However, in the nights of June 13th, 14th and July 1st of 2016, this object had the highest fluxes in VHE and during this period it had a EHBL-like behavior and the photohadronic model was unable to explain these VHE spectra with a single power-law. 

Recently, in a previous study, the HBL Mrk 501 had similar flaring events in 2005 and 2012 when the VHE events were also EHBL-like \citep{Sahu:2020tko}. For the first time, we used a two-zone photohadronic model and explained the VHE spectra of Mrk 501 very well. Here, we use the same two-zone model to explain the EHBL-like spectra of 1ES 1959+650. 
It is observed that, the low energy part of these spectra (zone-1) are explained very well using the standard photohadronic scenario where the spectral index has value $\delta_1=2.9$ which corresponds to high emission state. The high energy regime (zone-2) of the spectra are explained with $3.3\le \delta_2 \le 3.5$. We observed that the transition from zone-1 to zone-2 takes place around $E^{intd}_{\gamma} \simeq 1$ TeV
corresponding to seed photon energy $\epsilon_\gamma\sim 11.6$ MeV ($\sim 2.8\times 10^{21}\ \mathrm{Hz}$). As the proton spectral index $\alpha=2$ is the same in both the zones, this transition corresponds to changing of $\beta_1=0.9$ to $1.1\le \beta_2\le 1.5$ at $\epsilon_\gamma\sim 11.6$ MeV. This clearly shows that the transition from the HBL to EHBL is due to the change in the lower part of the tail region of the SSC spectrum, i.e., $\epsilon_\gamma < 11.6$ MeV.

An enhanced EBL contribution in the range $0.8-2\ \mathrm{\mu m}$ has been observed by CIBER and other experiments, setting an upper bound for the EBL SED of $\sim 28.7\ \mathrm{nW}\  \mathrm{n^{-2}}$ at $1.4\, \mathrm{\mu m}$. To account for this enhancement in the EBL contribution, we have increased the optical depth in the model of Franceschini et al. \citep{Franceschini:2008tp} up to 50\% and observed that by adjusting the normalization constants $F_1$ and $F_2$ we can still fit the observed VHE spectra well.

In the photohadronic model, the maximum proton energy during these flaring events is $E^{\textrm{max}}_p\simeq 84$ TeV corresponding to $E^{\textrm{max}}_{\gamma}\simeq 8.4$ TeV. Thus, the protons accelerated in the blazar jet during the above three nights are not energetic enough to be observed by cosmic ray detectors on Earth. Moreover, these protons can be deflected by the galactic magnetic field and will be difficult to correlate with the source.

The EHBL-like behavior of both 1ES 1959+650 and Mrk 501 are very similar in nature and in both cases it is the lower part of the tail region of the SSC spectrum which is modified during the EHBL-like behavior. As previous studies have established 1ES 1959+650 to be a HBL, this EHBL-like behavior seems to be transient in nature.
However, many more flaring events from different HBLs have to be studied to establish this fact.

We are thankful to Shigehiro Nagataki for many useful discussions. The work of S.S. is partially supported by DGAPA-UNAM (Mexico) Project No. IN103019 and partial support from CSU-Long Beach is gratefully acknowledged.

\bibliography{1es1959_650}{}
\bibliographystyle{aasjournal}
\end{document}